\documentclass[doublecol]{epl2} 
\usepackage{amssymb,graphicx,xspace,textcomp}
\usepackage[dvips]{hyperref}
\newcommand{\lao}{LaFeAsO\xspace}

\newcommand{\laf}{LaFeAsO$_{1-x}$F$_x$\xspace}
\newcommand{\smf}{SmFeAsO$_{1-x}$F$_x$\xspace}

\newcommand{\tc}{$T_c$\xspace}
\newcommand{\tn}{$T_N$\xspace}
\newcommand{\ts}{$T_S$\xspace}

\title{The intrinsic electronic phase diagram of iron-oxypnictide superconductors}

\shorttitle{The electronic phase diagram of iron-oxypnictide superconductors} 

\author{C. Hess\inst{1}\thanks{E-mail: \email{c.hess@ifw-dresden.de}} \and A. Kondrat\inst{1} \and A. Narduzzo\inst{1} \and J. E. Hamann-Borrero\inst{1} \and R. Klingeler\inst{1} \and J. Werner\inst{1} \and G. Behr\inst{1} \and B. B\"{u}chner\inst{1}}
\shortauthor{C. Hess \etal}

\institute{                    
  \inst{1} Leibniz-Institute for Solid State and Materials Research, IFW Dresden,
01171 Dresden, Germany, EU
}
\pacs{74.25.Dw}{Superconductivity phase diagrams}
\pacs{74.25.Fy}{Transport properties (electric and thermal conductivity, thermoelectric effects, etc.)}
\pacs{71.27.+a}{Strongly correlated electron systems; heavy fermions}

\abstract{
We present the first comprehensive derivation of the intrinsic electronic phase diagram of the iron-oxypnictide superconductors in the normal state based on the analysis of the electrical resistivity $\rho$ of both {\laf} and {\smf} for a wide range of doping. Our data give clear-cut evidence for unusual normal state properties in these new materials. 
In particular, the emergence of superconductivity at low doping levels is accompanied by distinct anomalous transport behavior in $\rho$ of the normal state which is reminiscent of the spin density wave (SDW) signature in the parent material.
At higher doping levels $\rho$ of {\laf} shows a clear transition from this pseudogap-like behavior to Fermi liquid-like behavior, mimicking the phase diagram of the cuprates. Moreover, our data reveal a correlation between the strength of the anomalous features and the stability of the superconducting phase. The pseudogap-like features become stronger in {\smf} where superconductivity is enhanced and vanish when superconductivity is reduced in the doping region with Fermi liquid-like behavior. 
}

\begin{document}

\maketitle
After the surprising discovery of iron pnictide superconductivity in {\laf} \cite{Kamihara2008},
strong evidence for unconventional superconductivity has rapidly emerged for this new material class.
Particularly striking is a close interplay between superconductivity and magnetism: A commensurate spin-density wave (SDW) ground state has been observed in the undoped parent compounds \cite{Cruz2008,Klauss2008,Drew2009,Kimber2008}, which is suppressed once superconductivity emerges upon doping \cite{Luetkens2008,Luetkens2009a,Drew2009,Drew2008}.
It is important to note that up to now {\laf} is the only pnictide system which exhibits a homogeneous superconducting state \cite{Luetkens2008}. Single crystals of intermetallic compounds {\cite{Aczel2008,Wang2009c,Ning2009a,Chu2009b,Lester2009}} and even stoichiometric materials where superconductivity is induced by external pressure \cite{Goko2008,Alireza2008} show a spatially inhomogeneous magnetic state coexisting with superconductivity. 
Thus {\laf} is the only known FeAs-based material which allows the study of the intrinsic electronic properties of superconducting species both in the superconducting and in the normal state.
 
%
%
%

From extensive work on other unconventional superconductors with a similar antiferromagnetic parent state such as heavy-fermion and cuprate superconductors it is known that the exploration of the normal state is
indispensable for the understanding of unconventional superconductivity. The electrical resistivity $\rho$ has been proven as a key experimental probe for this purpose \cite{Takagi1992,Ando2004,Custers2003,Gegenwart2008,Lohneysen2007}. 
In this letter we show that also the newly discovered pnictide superconductors show pronounced signatures of unusual normal state properties. 
In particular, the analysis of $\rho$ reveals distinct anomalies in the underdoped superconducting doping regime, which appear as remnants of the anomalies that accompany the structural and magnetic phase transitions of the non-superconducting parent compounds, and which strongly resemble pseudogap signatures of underdoped cuprate superconductors. Moreover, we find a transition from pseudogap-like to Fermi liquid-like behavior at further increased doping, and thus a strong resemblance to the cuprate phase diagram.

%

Polycrystalline {\laf} ($0\leq x\leq 0.2$) and {\smf} ($0\leq x\leq 0.1$) were prepared and characterized by powder X-ray diffraction (XRD) and wavelength-dispersive X-ray spectroscopy (WDX) \cite{Kondrat2008} and further characterized by magnetization \cite{Klingeler2008}, nuclear magnetic resonance \cite{Grafe2008}, and muon spin rotation ($\mu$SR) experiments \cite{Klauss2008,Luetkens2008,Luetkens2009a,Maeter2009}. 
The amount of impurity phases does not exceed the XRD resolution limit of $\sim5$\% except for slightly higher impurity concentrations at the respective maximal F-doping levels. {The largest peaks in the XRD spectra correspond to the insulating impurity phases LaOF, $\rm As_2O_3$ and $\rm Sm_2O_3$ and thus no significant effect on the transport properties is expected.}
According to the WDX analysis the ratio of the nominal and the actual F-content in the material is approximately 1:1 in {\laf} and 2.5:1 in {\smf}. All doping levels in the text thus correspond to the WDX result. 
Note, that for {\smf} the doping levels investigated by Liu et al. \cite{Liu2008a} and our nominal doping levels yield similar critical temperature values. 
We note further, that none of our superconducting samples shows a transition towards orthorhombic structure \cite{Kondrat2008,Hamann2008,Luetkens2008,Luetkens2009a}. The samples were investigated by four-probe $\rho$ measurements using an alternating DC-current. Details of the data analysis can be consulted at \href{http://xxx.lanl.gov/abs/0811.1601}{http://xxx.lanl.gov/abs/0811.1601}.

\begin{figure}
\includegraphics[width=\columnwidth,clip]{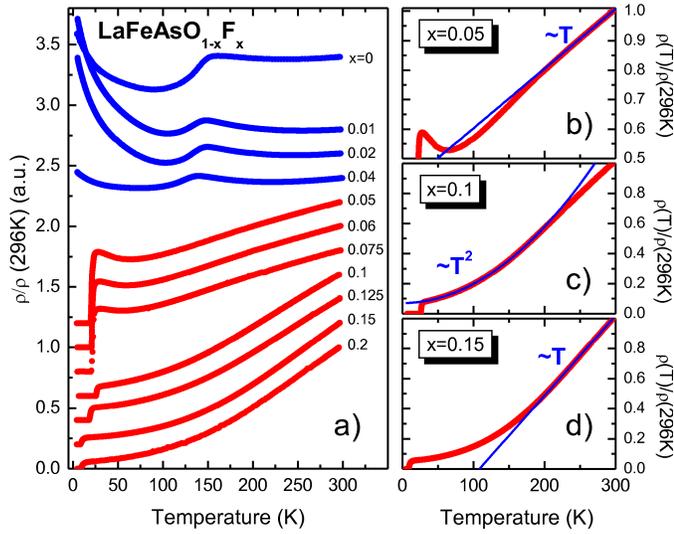}
\caption{a) $\rho(T)$ of {\laf}, normalized to $\rho(296~\rm K)$ and shifted for better visualization. Red (blue) data points refer to superconducting (non-superconducting) samples. 
b) Normalized $\rho(T)$ at $x=0.05$. The solid line shows the low-$T$ extrapolation of a linear fit between 250 and 300~K. c) Normalized $\rho(T)$ at $x=0.1$. The solid line represents a fit according to $\rho(T)=\rho_0+AT^2$ (fit between 50 and 150~K). d) Normalized $\rho(T)$ at $x=0.15$. The solid line shows the low-$T$ extrapolation of a linear fit between 250 and 300~K.
}
\label{fig:La_resistnorm}
\end{figure}

Fig.~\ref{fig:La_resistnorm}a) presents our data for $\rho$ of {\laf} as a function of temperature $T$ for all investigated doping levels $0\leq x\leq0.2$. Upon cooling, $\rho(T)$ of the non-superconducting parent compound {\lao} develops a deviation from the linear $T$-dependence near 300~K which leads to a maximum at $T_\mathrm{max}\sim 160$~K and a subsequent sharp drop with an inflection point at $T_\mathrm{drop}\sim 137$~K. We have previously shown \cite{Klauss2008} that the well established structural and SDW transitions at $T_S\approx T_\mathrm{max}$ and $T_N\approx T_\mathrm{drop}$ in this compound are intimately connected with these features.
Upon further cooling, $\rho(T)$ shows a minimum at $\sim70-90$~K followed by a strong low-$T$ upturn which is indicative of carrier localization, presumably arising from a SDW gap.
When increasing the F-doping level up to $x=0.04$, the essential features of $\rho(T)$ remain qualitatively the same as in the undoped material. In particular, $T_\mathrm{max}$ and $T_\mathrm{drop}$ are slightly shifted towards lower $T$, and a little broadening of the corresponding anomalies occurs.

A further increase of $x$ leads to a sudden occurrence of superconductivity with rather high critical temperature \tc \cite{Luetkens2009a} and drastic changes in $\rho(T)$ in the normal state. For $0.05\leq x\leq 0.075$, a low-$T$ upturn ($T\lesssim 60$~K) is still present before entering the superconducting state, which is reminiscent of the low-$T$ upturn of the low-doping compounds. At high $T$, however, the clear features at $\sim150$~K of the non superconducting samples have disappeared. Instead, $\rho$ increases monotonically for $T\gtrsim60$~K up to 300~K.
A close inspection of this increase reveals a surprising feature: while $\rho(T)$ becomes linear at $T\gtrsim250$~K, it drops below the low-$T$ extrapolation of this linearity (cf. fig.~\ref{fig:La_resistnorm}b for a representative example). This drop is connected with an inflection point at $T_\mathrm{drop}\approx 150$~K, which can be conveniently extracted from the derivatives $d\rho/dT$. Thus, despite the suppression of the actual structural and magnetic transitions, a distinct reminiscent feature with the same energy scale persists in these superconducting samples. 

Yet another drastic systematic change of $\rho(T)$ is observed when the doping level enters the regime $0.1\leq x\leq0.2$. Here, instead of the low-$T$ upturn, we find $\rho(T)=\rho_0+AT^2$ ($\rho_0=\mathrm{const}$) from just above \tc up to $\sim200$~K, i.e., a Fermi-liquid-like behavior \cite{Sefat2008} which indicates enhanced electron-electron interaction (cf. fig.~\ref{fig:La_resistnorm}c for a representative example). A maximum $T_c=26.8$~K is observed at $x=0.1$ which quickly diminishes with further increasing $x$ \cite{Luetkens2009a}. Simultaneously, the anomaly connected with the inflection point becomes weaker at higher $T_\mathrm{drop}$ and eventually vanishes completely for $x\geq0.15$ where the quadratic low-$T$ increase shows a smooth crossover to a linear high-$T$ behavior (cf. fig.~\ref{fig:La_resistnorm}d).

At this point, a striking similarity to the electronic phase diagram of hole doped cuprates becomes apparent. In the underdoped region $0.05\leq x\leq0.075$, the overall temperature dependence, i.e., the low-$T$ upturn together with the feature at 150~K and the linear increase at higher $T$, almost perfectly mimics the resistivity of underdoped cuprate superconductors \cite{Takagi1992}. In particular, the feature at $T_\mathrm{drop}$ amazingly resembles well known pseudogap signatures \cite{Takagi1992,Ando2004}. The finding of a Fermi-liquid-like behavior at higher doping levels $x\geq0.1$ (hereafter called overdoped region) is a further similarity if compared with the normal state of overdoped cuprates where a qualitatively similar $\rho(T)$ is observed \cite{Takagi1992,Ando2004,Nakamae2003}.

\begin{figure}
\includegraphics[width=\columnwidth,clip]{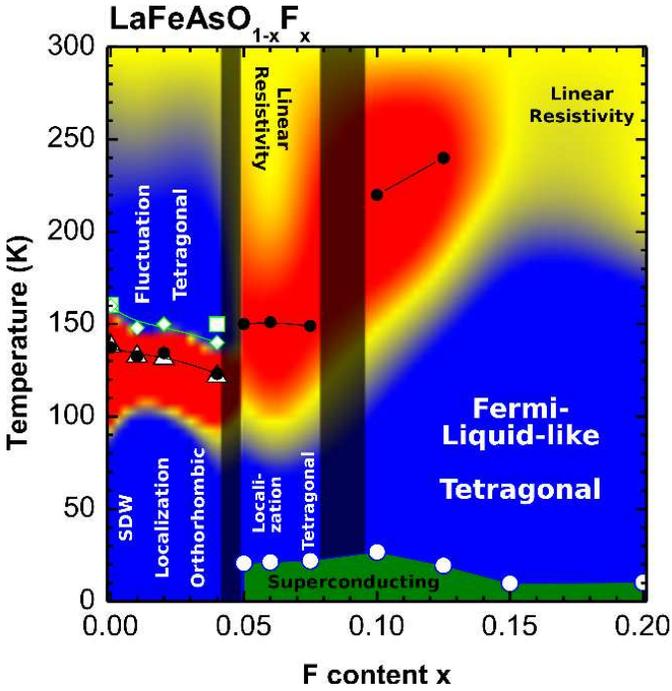}
\caption{Phase diagram of {\laf} as a function of the doping level $x$ and temperature $T$, highlighting the unusual $\rho(T)$ as compared to that of the (approximately) linear $\rho(T)$ near 300~K.
The latter ($\frac{d\tilde \rho}{dT}(T)=\frac{d\rho}{dT}(T)-\frac{d\rho}{dT}(\rm 296~K)\approx0$) gives rise to the yellow areas. The blue regions ($d\tilde \rho/dT<0$) indicate carrier localization/fluctuation and Fermi liquid-like behavior for $x\leq0.075$ and $x\geq0.1$, respectively. Across the whole phase diagram the red areas ($d\tilde \rho/dT>0$) are centered around $T_\mathrm{drop}$ and mark the signatures of the structural/magnetic transitions ($x\leq0.04$) and the corresponding remnant feature ($x\geq0.05$). The dark bars separate the non-superconducting, the underdoped and the overdoped superconducting regimes. The diagram shows also data points for \tc (\textbigcircle), $T_\mathrm{drop}$ (\textbullet), $T_\mathrm{max}$ ($\Diamond$), and, where available, \tn ($\triangle$) and \ts ($\square$) from $\mu$SR and XRD experiments \cite{Luetkens2009a}. }
\label{fig:La_phasendiagramm}
\end{figure}

We summarize our major findings for {\laf} in the phase diagram shown in fig.~\ref{fig:La_phasendiagramm}.
In the underdoped superconducting region the signatures of both the resistivity drop and the low-$T$ localization clearly 'survive' despite the suppression of the structural/magnetic transitions and the occurrence of superconductivity. These anomalous features strongly suggest that fluctuations connected to the SDW are still present. They apparently lead to a renormalization of the charge carriers, thus playing a major role in the physics of the superconductivity in the system. The observation of $T_\mathrm{drop}$ in the underdoped region being about the same as at $x\leq0.04$ shows that these fluctuations are of a similar energy scale as the actual SDW state. 
Intriguingly, in the overdoped region the fluctuation features vanish and Fermi liquid-like behavior becomes increasingly dominating over a large $T$-range. In view of this strong change towards less unconventional normal state properties and the similarity to the cuprate phase diagram quantum critical behavior should be considered in this material \cite{Lee2006,Vojta2003}.

\begin{figure}
\includegraphics[width=\columnwidth,clip]{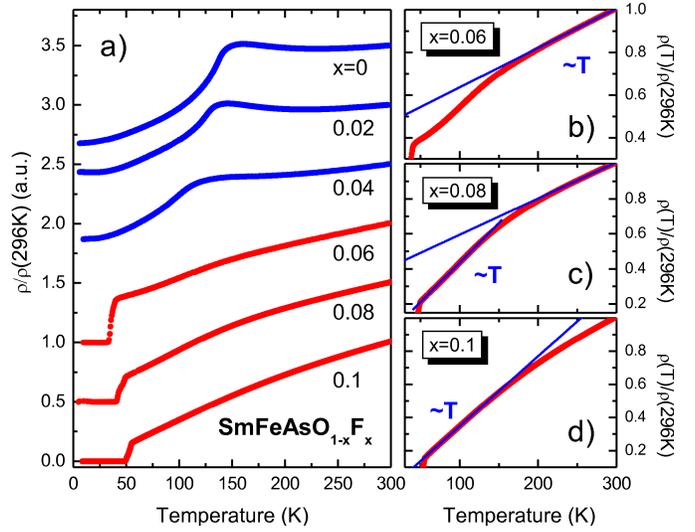}
\caption{a) $\rho(T)$ {\smf}, normalized to $\rho(296~\rm K)$ and shifted for better visualization. Red (blue) data points refer to superconducting (non-superconducting) samples with $T_c=36.2$~K, 44.5~K, 52.1~K at $x=0.06$, 0.08, 0.1. 
b) Normalized $\rho(T)$ at $x=0.06$. The solid line shows the low-$T$ extrapolation of a linear fit between 250 and 300~K. c) Normalized $\rho(T)$ at $x=0.08$. Solid lines represent linear fits to the low-$T$ ($50-120$~K) and high-$T$ ($250-300$~K) linear regimes. d) Normalized $\rho(T)$ at $x=0.1$. The solid line shows a linear fit to the region $60-130$~K.}
\label{fig:Sm_resistnorm}
\end{figure}
We now turn to $\rho(T)$ of {\smf}, which cover the doping range $0\leq x\leq0.1$ (see fig.~\ref{fig:Sm_resistnorm}). The direct comparison with {\laf} reveals extensive similarities, in particular the structural/magnetic transitions \cite{Maeter2009,Hamann2008} which are observed in the non-superconducting compounds lead to the same salient features. Interestingly, the pseudogap-like anomalies become even more pronounced and noticeable without further analysis.

\begin{figure}
\includegraphics[width=1\columnwidth,clip]{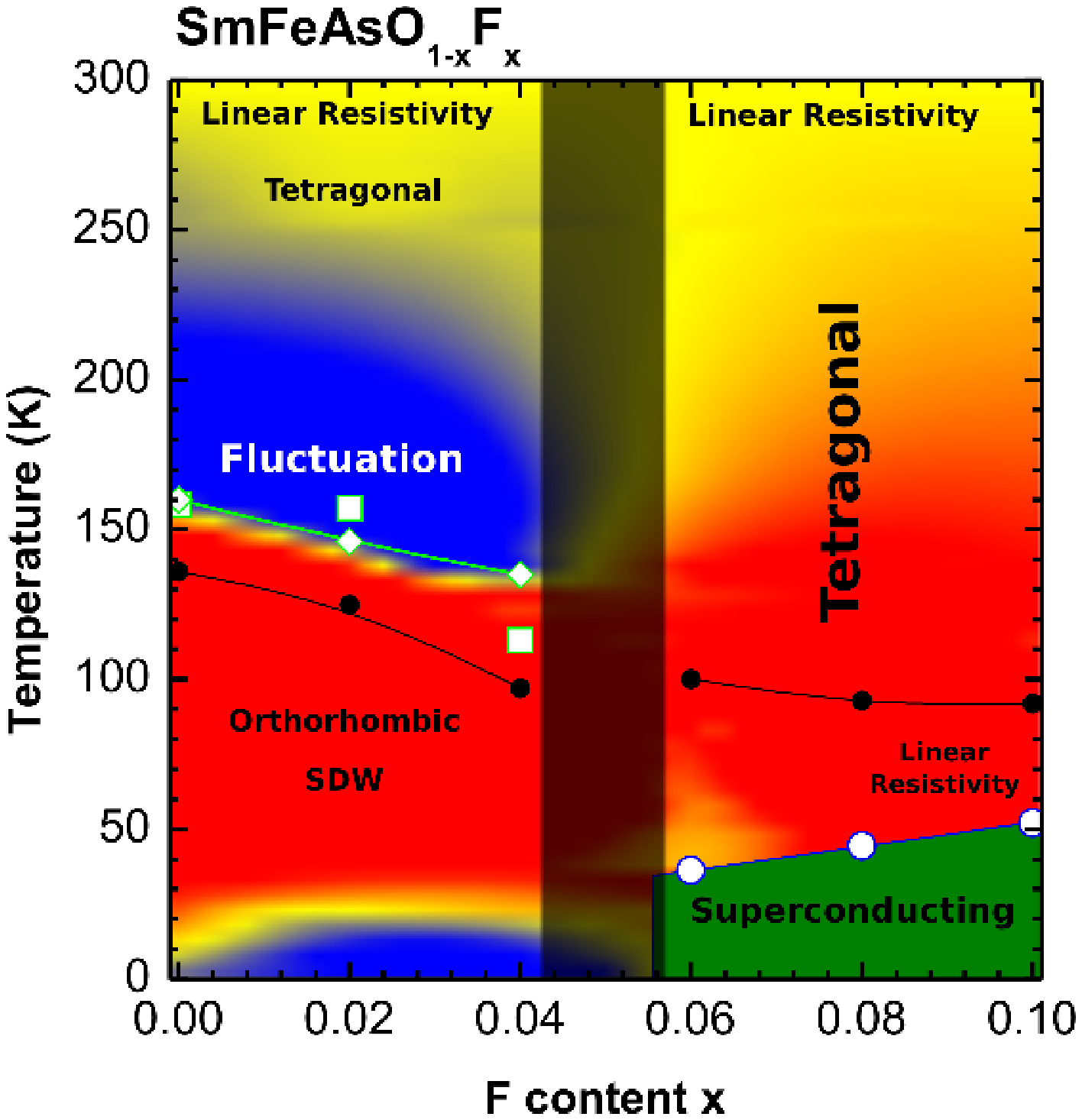}
\caption{Phase diagram of {\smf}. As in fig.~\ref{fig:La_phasendiagramm}, the diagram highlights where $d\tilde \rho/dT$ is larger (red), smaller (blue) or equal to the slope near 300~K (yellow). 
The dark bar marks the transition from the non-superconducting to the superconducting region.
The diagram shows also data points for \tc (\textbigcircle), $T_\mathrm{drop}$ (\textbullet), $T_\mathrm{max}$ ($\Diamond$) extracted from the resistivity, and \ts ($\square$) from XRD experiments \cite{Hamann2008}. Note, that at $x=0.04$ $T_\mathrm{max}$ has been extracted after subtracting a linear background which matches $\rho(T)$ near 300~K.}
\label{fig:Sm_phasendiagramm}
\end{figure}

The doping evolution of the electronic properties of {\smf} is summarized in the phase diagram in fig.~\ref{fig:Sm_phasendiagramm} which apparently corresponds very well to that of {\laf} in the non-superconducting and underdoped region. Note, that a Fermi liquid-like region is absent for the investigated doping levels. We further point out that a low-$T$ upturn as is found in $\rm LaFeAsO_{1-x}F_x$ is absent in the Sm-based compounds for all studied samples. Apparently, the presence of the Sm moments modifies the electronic structure in such a manner that the tendency towards localizing behavior is reduced, which indicates a significantly reduced SDW gap as compared to $\rm LaFeAsO_{1-x}F_x$. This conjecture is consistent with recent Hall-effect data \cite{Liu2008a,McGuire2008} which yield a much stronger SDW-induced reduction of charge carriers for the La-based system.

One might speculate that the reinforcement of the pseudogap-like anomalies seen for the superconducting samples is related to substantial remnants of the structural and magnetic transitions at low doping. In fact, recent $\mu$SR data \cite{Drew2009,Drew2008} provide clear-cut evidence for magnetic order and/or slow magnetic fluctuation in all superconducting species of {\smf}, and thus corroborate this notion.

The direct comparison of the phase diagrams shown in Figures~\ref{fig:La_phasendiagramm} and \ref{fig:Sm_phasendiagramm} allows important conclusions about the correlation of anomalous transport behavior and superconductivity.
Interestingly, \tc is strongly enhanced in {\smf} despite the reinforcement of the pseudogap-like features and static magnetism \cite{Drew2009,Drew2008}, and thus suggests the intimate connection between these phenomena. On the other hand, in {\laf} the fading of the pseudogap-like anomalies at high doping levels is accompanied by a weakening of the superconducting state which is readily seen by the strongly reduced $T_c$. Thus a key approach to unravel the nature of superconductivity in the iron pnictide materials is the understanding of the nature of the electronic renormalization which is connected with the fluctuation of the SDW state.

\begin{figure}
\includegraphics[width=1\columnwidth,clip]{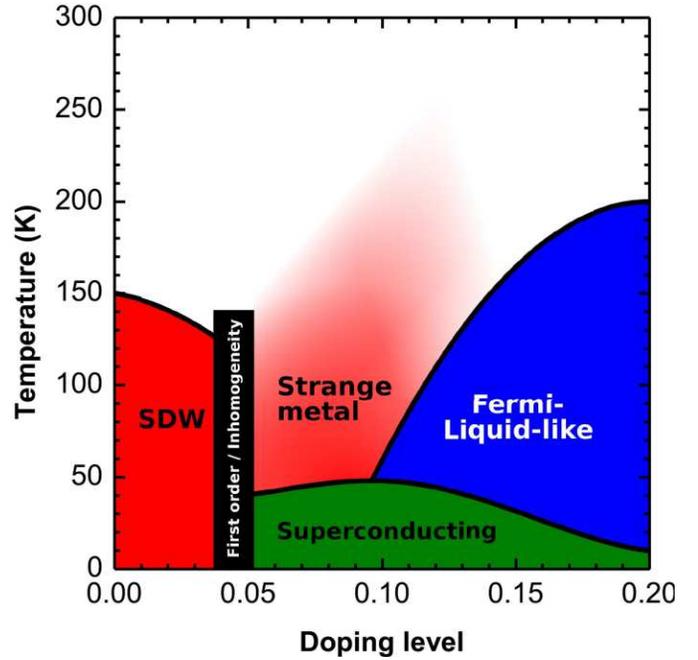}
\caption{Schematic phase diagram of FeAs superconductors. The red, green and blue regions indicate the SDW, superconducting and Fermi liquid-like regions, respectively. In the 'strange-metal' region  unusual $\rho(T)$ is observed.  The black bar marks the doping-driven transition from the SDW state to superconductivity, which is of first order character and/or accompanied by strong inhomogeneity, i.e., the coexistence of the superconductivity and static magnetism.}
\label{fig:phdschem_fig}
\end{figure}

To conclude, our data unambiguously show that the high temperature superconductivity in the iron pnictides is intimately correlated with anomalous transport properties in the normal state. This is clearly seen in the evolution of the superconducting dome as a function of doping, which is accompanied by strong changes of the electrical resistivity at higher $T$. In particular, a rather conventional Fermi liquid-like behavior occurs in the overdoped region where the critical temperature \tc is small. {True high-$T_c$ superconductivity with $T_c> 20$~K} is found only in the lower doped region of the phase diagram where the resistivity shows pronounced anomalies which are clearly related to the anomalous SDW state of the undoped parent compounds. Note, that the stronger manifestation of these anomalies in {\smf} is connected with an enhancement of \tc.
Summing up all these anomalous features, the electronic phase diagram of the FeAs superconductors (cf. fig.~\ref{fig:phdschem_fig}) yields a striking resemblance to the generic phase diagram of cuprate superconductors \cite{Vojta2000,Vojta2003} and that of other unconventional superconductors in the vicinity of a quantum critical point \cite{Gegenwart2008,Lohneysen2007}. However, unlike the latter examples, the doping-driven transition from the non-superconducting magnetic ground state to superconductivity appears to be first order-like and/or accompanied by inhomogeneity \cite{Drew2008,Drew2009,Goko2008,Aczel2008}.

%
%
We thank S.-L. Drechsler, J. Geck, H.-J. Grafe, H.-H. Klauss and A. U. B Wolter and for valuable discussions and M. Deutschmann, S.
M\"uller-Litvanyi, R. M\"uller, and S. Ga{\ss} for technical support. This work has been supported by the
Deutsche Forschungsgemeinschaft, through BE1749/12 and through FOR 538 (BU887/4).

\begin{figure}[h]
\includegraphics[width=1\columnwidth,clip]{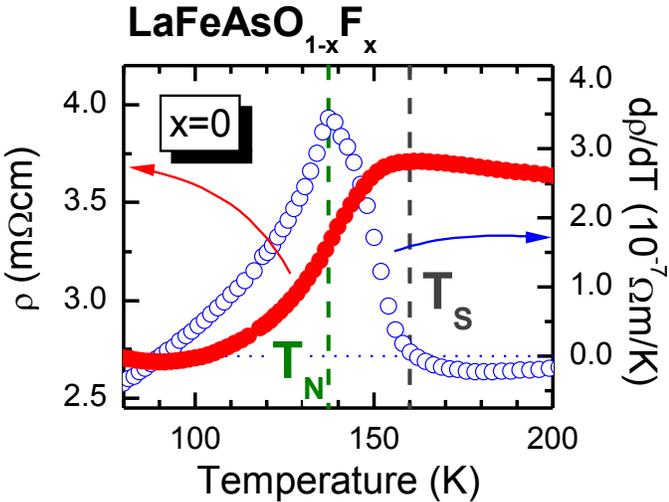}
\caption{$\rho(T)$ and the derivative $d\rho/dT$ at $x=0$ near \tn and \ts. Vertical dashed lines mark \tn and \ts from $\mu$SR and x-ray diffraction experiments, respectively \cite{Luetkens2009a}. The blue dotted line indicates the zero of the $d\rho/dT$ axis.}
\label{fig:La_illustration}
\end{figure}

\begin{figure}[h]
\includegraphics[width=0.8\columnwidth,clip]{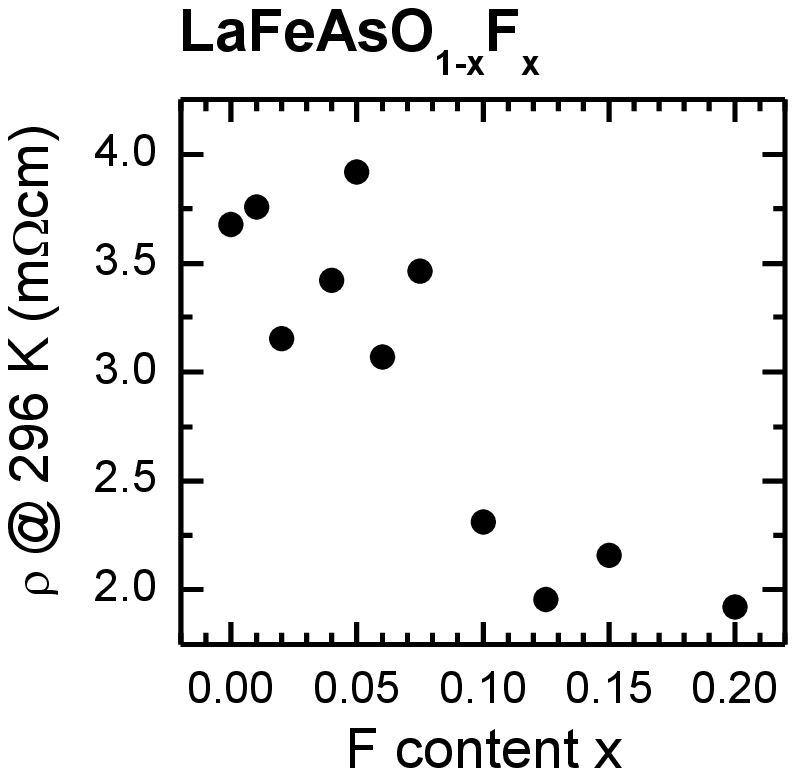}
\caption{$\rho$ of \laf at 296~K as a function of doping. The geometrical error amounts to less than 15\%.}
\label{fig:La_absolute}
\end{figure}

\begin{figure}[h]
\includegraphics[width=1\columnwidth,clip]{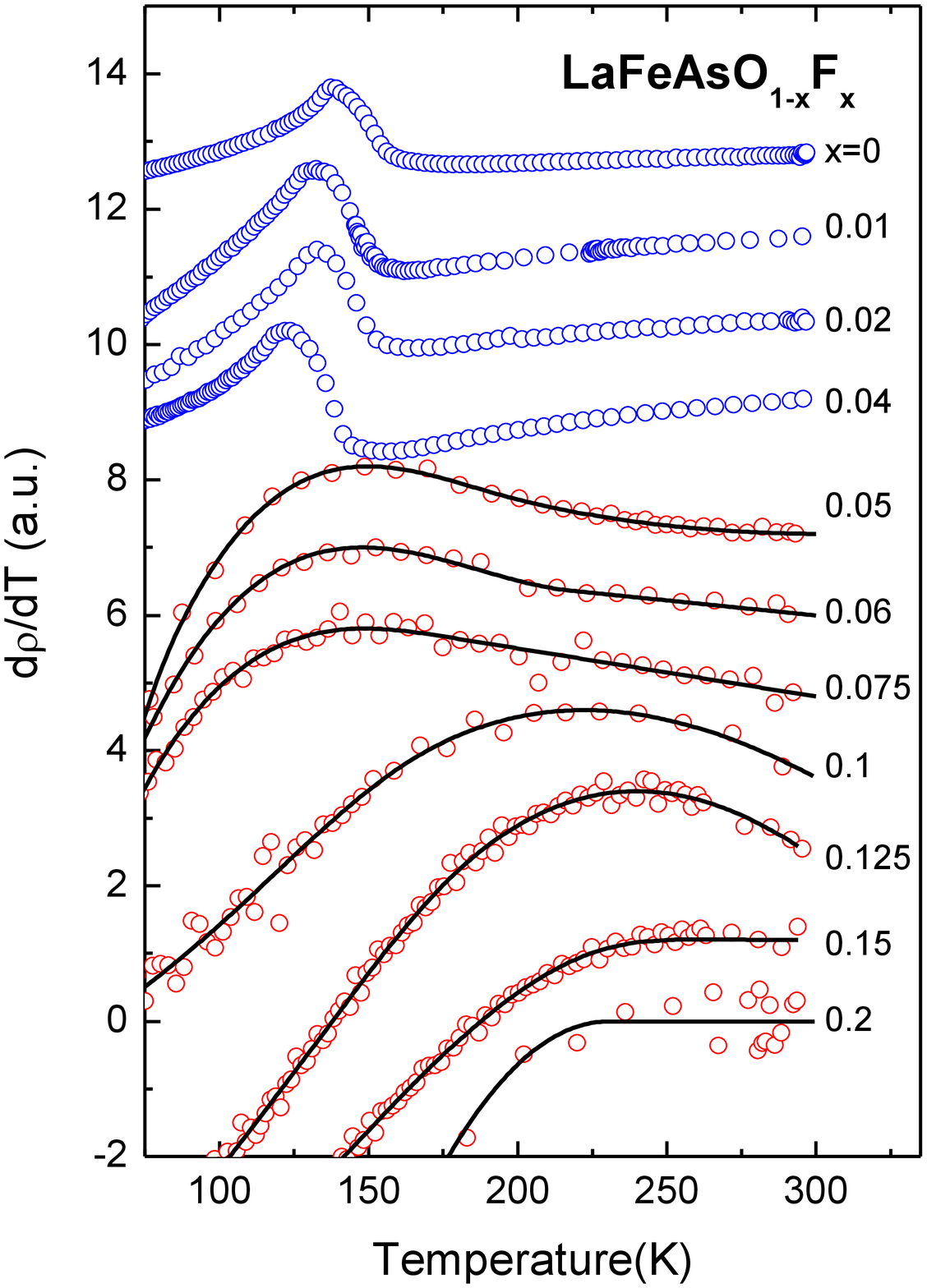}
\caption{$d\rho/dT$ of \laf for all studied doping levels $x$ as a function of temperature. The shown data have been multiplied by suitable factors and shifted for better visualization. Red (blue) data points refer to superconducting (non-superconducting) samples. The solid lines are guides to the eye.}
\label{fig:La_ableitungen}
\end{figure}

\begin{figure}[h]
\includegraphics[width=1\columnwidth,clip]{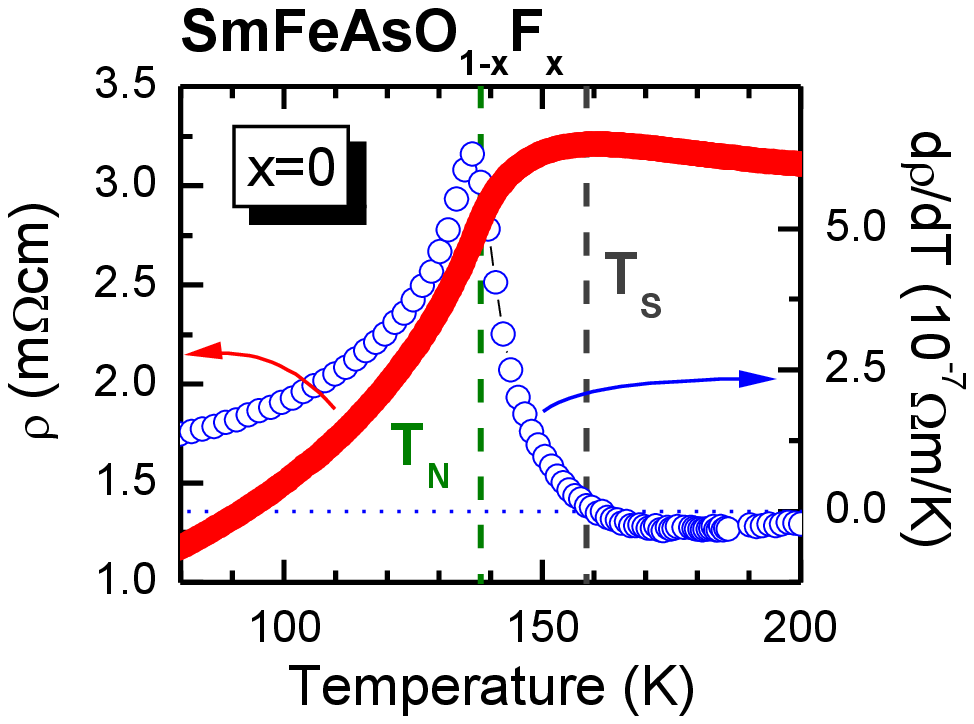}
\caption{$\rho(T)$ and the derivative $d\rho/dT$ at $x=0$ near \tn and \ts. Vertical dashed lines mark \tn and \ts from $\mu$SR \cite{Maeter2009} and x-ray diffraction \cite{Hamann2008} experiments, respectively. The blue dotted line indicates the zero of the $d\rho/dT$ axis. }
\label{fig:Sm_illustration}
\end{figure}

\begin{figure}[h]
\includegraphics[width=0.8\columnwidth,clip]{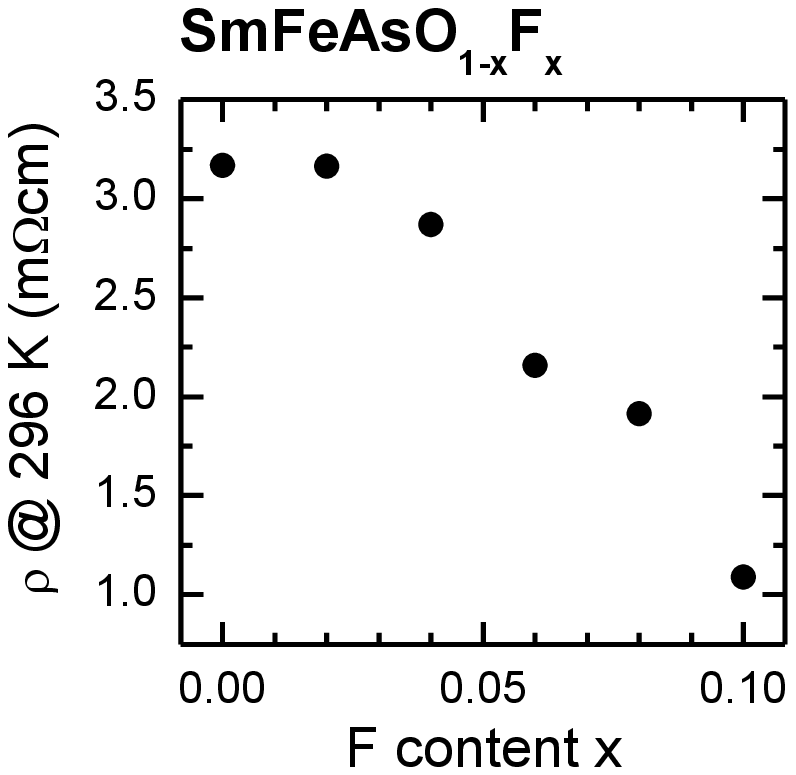}
\caption{$\rho$ of \smf at 296~K as a function of doping. The geometrical error amounts to less than 15\%.}
\label{fig:Sm_absolute}
\end{figure}

\begin{figure}[h]
\includegraphics[width=1\columnwidth,clip]{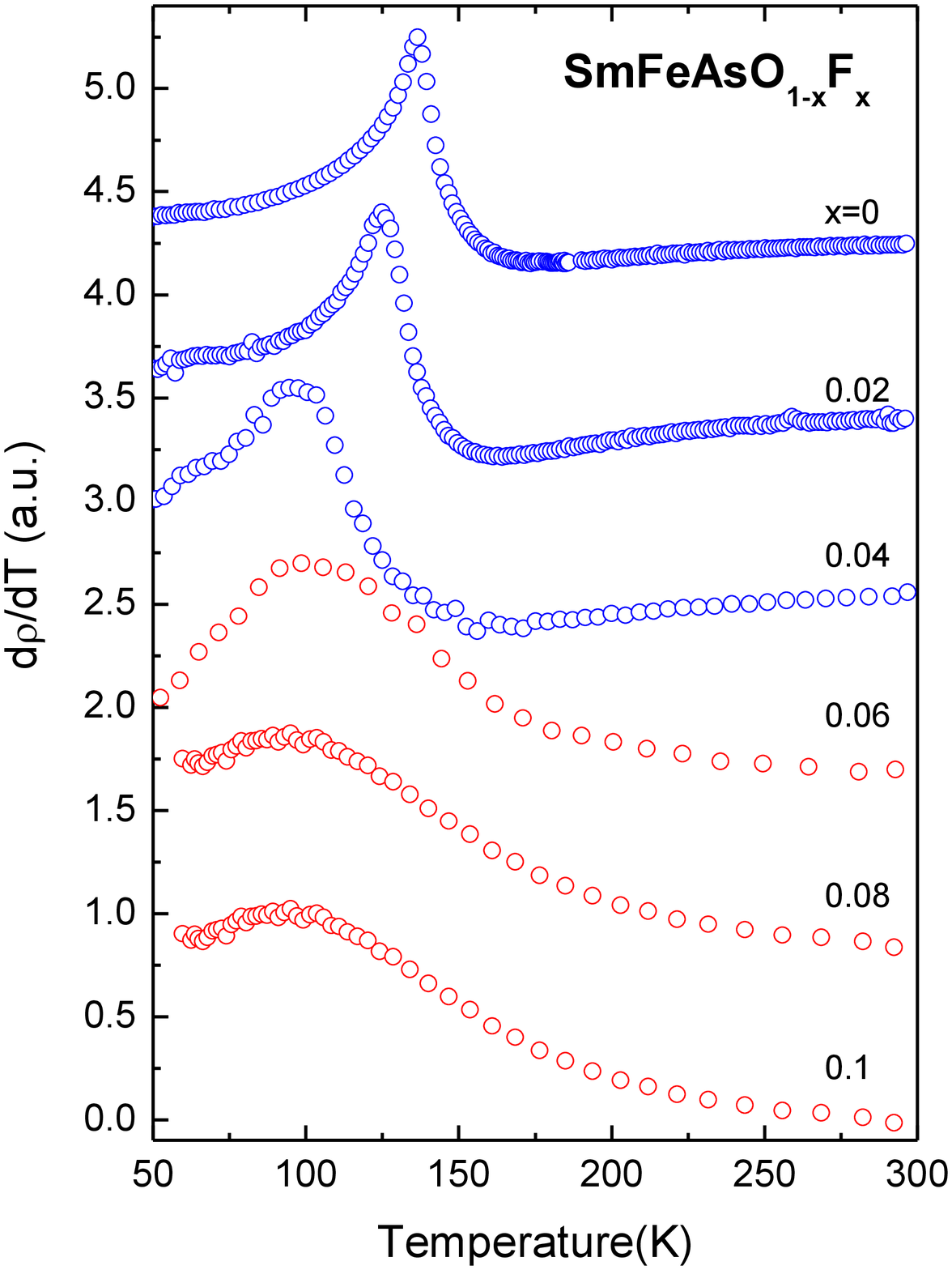}
\caption{$d\rho/dT$ of \smf for all studied doping levels $x$ as a function of temperature. The shown data have been multiplied by suitable factors and shifted for better visualization. Red (blue) data points refer to superconducting (non-superconducting) samples.}
\label{fig:Sm_ableitungen}
\end{figure}

\textbf{Supplementary information} -- 
The following figures illustrate specific features of $\rho(T)$ of \laf and \smf. Supplementary Figure~\ref{fig:La_illustration} shows how $T_S$ and $T_N$ can be inferred from the resistivity of the non-superconducting samples of \laf. The doping trend of the absolute value of $\rho$ of \laf at room temperature is given in Figure~\ref{fig:La_absolute}. The derivatives $d\rho/dT$ for all doping levels of \laf are shown in Figure~\ref{fig:La_ableitungen}. These data were used to extract $T_\mathrm{drop}$ and $T_\mathrm{max}$. In order to highlight the unusual normal state properties we compare $\frac{d\rho}{dT}(T)$ with its value at room temperature where $\frac{d\rho}{dT}\approx \mathrm{const}$. For this purpose, we define $\frac{d\tilde \rho}{dT}(T)=\frac{\rho}{dT}(T)-\frac{d\rho}{dT}(296~K)$ for the use in Figures~2 and 4 of the article. Figures~\ref{fig:Sm_illustration}, \ref{fig:Sm_absolute} and \ref{fig:Sm_ableitungen} show the respective data for \smf.

\newpage


\end{document}